\pdfoutput=1
\documentclass[a4paper,11pt]{article}
\usepackage[utf8]{inputenc}
\usepackage[sort&compress,numbers,colon]{natbib}
\usepackage{booktabs}
\usepackage{amsmath}
\usepackage{amssymb}
\usepackage{amsfonts}

\usepackage{bbold}
\usepackage{bm}

\newcommand{\cHPS}{\cite{Harrison:2002er,Harrison:2002kp}}
\newcommand{\cAfour}{\cite{Ma:2001lr,Babu:2003fk,Ma:2004qy,Babu:2005uq,Altarelli:2005uq,He:2006dk,Altarelli:2006qy}}

\usepackage{a4wide}
\usepackage{graphicx}
\usepackage{url}
\usepackage{mathrsfs}
\usepackage[hang,normal,footnotesize]{caption}
\usepackage{color}
\usepackage[colorlinks]{hyperref}
\usepackage{bbm}
\usepackage{units}

\hypersetup{
pdfauthor={Martin Holthausen, Kher Sham Lim, Manfred Lindner},
pdftitle={Lepton Mixing Patterns from a Scan of Finite Discrete Groups}
}

\makeatletter
\renewcommand{\fnum@table}{\textbf{\tablename~\thetable}}
\renewcommand{\fnum@figure}{\textbf{\figurename~\thefigure}}
\makeatother

\newcommand{\diag}{\ensuremath{\mathrm{diag}}}
\newcommand{\I}{\ensuremath{\mathrm{i}}}

\newcommand{\eV}{\ensuremath{\,\mathrm{eV}}}

\newcommand{\braket}[1]{\ensuremath{\left<#1\right>}}
\newcommand{\ev}[1]{\ensuremath{\left\langle#1\right\rangle}}

\newcommand{\ra}[1]{\renewcommand{\arraystretch}{#1}}

\newcommand{\be}{\begin{equation}}
\newcommand{\ee}{\end{equation}}
\newcommand{\ba}{\begin{eqnarray}}
\newcommand{\ea}{\end{eqnarray}}

\renewcommand{\subsubsection}[1]{\vspace{1ex}\mathversion{bold}{\bf #1:}\mathversion{normal}}

\newcommand{\GAP}{\texttt{GAP}}

\newcommand{\vev}[1]{\ensuremath{\left\langle#1\right\rangle}}
\newcommand{\SmallGroups}{\texttt{SmallGroups}}

\newcommand{\abs}[1]{\ensuremath{\left\vert#1\right\vert}}

\begin{document}

\allowdisplaybreaks[1]


\begin{titlepage}

\begin{center}
{\Huge\sffamily\bfseries 
Lepton Mixing Patterns from a Scan of Finite Discrete Groups
}
\\[10mm]
{\large
Martin Holthausen\footnote{\texttt{martin.holthausen@mpi-hd.mpg.de}}, Kher Sham Lim\footnote{\texttt{khersham.lim@mpi-hd.mpg.de}} and Manfred Lindner\footnote{\texttt{lindner@mpi-hd.mpg.de}}}
\\[5mm]
{\small\textit{
Max-Planck Institut f\"ur Kernphysik, Saupfercheckweg 1, 69117
Heidelberg, Germany
}}

\end{center}
\vspace*{1.0cm}
\date{\today}

\begin{abstract}
\noindent
The recent discovery of a non-zero value of the mixing angle $\theta_{13}$ has ruled out tri-bimaximal mixing as the correct lepton mixing pattern generated by some discrete flavor symmetry (barring large next-to-leading order corrections in concrete models). In this work we assume that neutrinos are Majorana particles and perform a general scan of all finite discrete groups with order less than 1536 to obtain their predictions for lepton mixing angles. To our surprise, the scan of over one million groups only yields 3 interesting groups that give lepton mixing patterns which lie within 3-sigma of the current best global fit values. A systematic way to categorize such groups and the implications for flavor symmetry are discussed. 
\end{abstract}

\end{titlepage}

\pagestyle{plain}
\section{Introduction}
The origin of flavor is one of the important questions of beyond the Standard Model physics. All entries of the lepton mixing matrix, or better known as the Pontecorvo-Maki-Nakagawa-Sakata (PMNS) matrix, are of order one, with the exception of $U_{e3}$. Compared to the Cabibbo-Kobayashi-Maskawa (CKM) matrix whose off-diagonal entries are small, the very different form of the PMNS matrix seems to suggest a different origin of the two matrices. One popular approach to the flavor puzzle is to invoke (spontaneously broken) symmetries to describe the observed patterns. The leptonic mixing angles can be determined solely from flavor symmetry considerations (up to permutations of rows and columns of the mixing matrix). This is possible if the charged lepton and neutrino mass matrices exhibit the misaligned remnant symmetries under which charged leptons and neutrinos transform as three inequivalent singlets, as will be reviewed in the next chapter. 

Assuming the remnant symmetries to be part of the original symmetry group (and not a result of an accidental symmetry) one can then determine mixing patterns from the structure of discrete symmetry groups. For review on discrete flavor symmetries and their application in model building see \cite{Altarelli:2010qy,Ishimori:2010zr,Grimus:2012dk}. For example the symmetry group $A_4$ \cAfour~ and $S_4$ \cite{Lam:2007fk,Lam:2008fj,Lam:2008sh} can lead to the tri-bimaximal mixing pattern (TBM) by Harrison, Perkins and Scott~\cHPS. With the latest global fits results \cite{Fogli:2012ua,Tortola:2012te,GonzalezGarcia:2012sz} (see Table \ref{tab:osc-parameters}) on the non-zero mixing angle $\theta_{13}$ measured by DAYA BAY \cite{An:2012eh}, RENO \cite{Ahn:2012nd} and DOUBLE CHOOZ \cite{Abe:2011fz}, TBM is ruled out and one is prompted to rethink the approach to lepton flavor based on discrete groups. One possibility is to build models which lead to TBM on leading order and allow for large next-to-leading order (NLO) corrections. Here the problem 
usually is that quite often there are many different NLO corrections, which limits the predictivity of the models.  Another approach is to look for new groups that predict a different type of leptonic mixing pattern i.e. a new starting point about which models could be built. In this paper, we shall follow the second route and therefore perform a scan of all possible finite discrete groups of the order less than 1536 with the help of the computer algebra program \GAP~\cite{GAP4:2011,REPSN:2011,SmallGroups:2011,SONATA:2003}. To our surprise, only three finite discrete groups can yield the neutrino mixing angles allowed by the experimental constraints. 

This paper is organized as follows: in Section \ref{sec:lepton-mixing-finite-groups} we will present the group theoretical procedure to obtain the PMNS matrix from a finite symmetry group. This section might be skipped by readers familiar with the methodology. The method of scanning through all the groups of order less than 1536 and the relevant results are presented in Section \ref{sec:new-starting-points} and finally we conclude in Section \ref{sec:conclusion}.

\begin{table}[ht]
\centering
\ra{1.3}
\begin{tabular}{lcccccc}\toprule
  & $\Delta m_{21}^2$& $\abs{\Delta m_{31}^2}$ &  $\sin^2 \theta_{12}$ &  $\sin^2 \theta_{23}$ &  $\sin^2 \theta_{13}$ &  $\delta$ \\
  & $[10^{-5}\eV^2]$& $[10^{-3}\eV^2]$ &  $[10^{-1}]$ & $[10^{-1}]$&$[10^{-2}]$   &  $[\pi]$ \\   \cmidrule{2 -7} 
 best fit & $7.62^{+ .19}_{- .19} $& $2.55^{+ .06}_{- .09} $ &  $3.20^{+ .16}_{- .17} $ &  $6.13^{+ .22}_{- .40}$ &  $2.46^{+ .29}_{- .28} $ &  $0.8^{+1.2 }_{-.8 }  $ \\ 
 $3 \sigma$ range& $7.12-8.20 $& $2.31-2.74 $ &  $2.7-3.7$ &  $3.6-6.8 $ &  $1.7-3.3 $ &  $0-2  $ \\ 
\bottomrule
\end{tabular} 
\caption{Global fit of neutrino oscillation parameters (for normal ordering of neutrino masses) adapted from~\cite{Tortola:2012te}. The errors of the best fit values indicate the one sigma ranges. In the global fit there are two nearly degenerate minima at $\sin^2 \theta_{23}=0.430^{+ .031}_{- .030}$, see Figure \ref{fig:para-plot}.}
\label{tab:osc-parameters}
\end{table}
\ra{1}

\section{Leptonic Mixing from Remnant Symmetries}
\label{sec:lepton-mixing-finite-groups}
Lepton mixing can be derived from a flavor symmetry breaking to remnant symmetries in the charged lepton and neutrino masses respectively. In concrete models, this is usually achieved via a spontaneous breaking using some scalar fields charged under this symmetry into different directions of flavor space. The charge assignments are chosen such that there are different residual symmetries in the charged lepton and neutrino sectors. The misalignment between the two residual symmetries generates the PMNS matrix \cite{Lam:2007fk,Lam:2008sh,Lam:2008fj,Toorop:2011jn,de-Adelhart-Toorop:2012fv,Lam:2012he}. In this method, only the structure of flavor symmetry group and its remnant symmetries are assumed and we do not consider the breaking mechanism i.e. how the required vacuum alignment needed to achieve the remnant symmetries is dynamically realized.

The PMNS matrix is defined as
\begin{align}
U_{\mathrm{PMNS}}=V_e^\dagger V_\nu
\end{align}
and can be determined from the unitary matrices $V_e$ and $V_\nu$ satisfying
\begin{align}
V_e^T M_e M_e^\dagger V_e^*=\diag(m_e^2,m_\mu^2,m_\tau^2)\qquad \mathrm{and} \qquad 
V_\nu^T M_\nu V_\nu=\diag(m_1, m_2, m_3),
\end{align} 
where the mass matrices are defined by $\mathcal{L}=e^T M_e e^c+ \frac{1}{2}\nu^T M_\nu \nu.$ We will now review how certain mixing patterns can be understood as a consequence of mismatched horizontal symmetries acting on the charged lepton and neutrino sectors \cite{Lam:2007fk,Lam:2008sh,Lam:2008fj,Toorop:2011jn,de-Adelhart-Toorop:2012fv,Lam:2012he}\footnote{We here follow the presentation and convention in 
\cite{Toorop:2011jn,de-Adelhart-Toorop:2012fv}.}. Let us assume for this purpose that there is a (discrete) symmetry group $G_f$ under which the left-handed lepton doublets $L=(\nu,e)^T$ transform under a faithful unitary 3-dimensional representation $\rho:G_f\rightarrow GL(3,\mathbb{C})$:
\begin{align}
L\rightarrow \rho(g) L, \qquad g \in G_f.  
\end{align}
The experimental data clearly shows \emph{(i)} that all lepton masses are unequal and \emph{(ii)} there is mixing amongst all three mass eigenstates. Therefore this symmetry cannot be a symmetry of the entire Lagrangian but it has to be broken to different subgroups $G_e$ and $G_\nu$ (with trivial intersection) in the charged lepton and neutrino sectors, respectively. If the fermions transform as 
\begin{align}
e\rightarrow \rho(g_e) e, \qquad \nu \rightarrow \rho(g_\nu) \nu , \qquad g_e \in G_e, g_\nu \in G_\nu,
\end{align}
for the symmetry to hold, the mass matrices have to fulfil
\begin{align}
\rho(g_e)^T M_e M_e^\dagger \rho(g_e)^*=M_e M_e^\dagger \qquad \mathrm{and} \qquad\rho(g_\nu)^T M_\nu \rho(g_\nu)=M_\nu.
\label{eq:symmcond}
\end{align}
Choosing $G_e$ or $G_\nu$ to be a non-abelian group would lead to a degenerate mass spectrum, as their representations cannot be decomposed into three inequivalent 1-dimensional representations of $G_e$ or $G_\nu$. This scenario is not compatible with the case of three distinguished neutrino and charged lepton masses and we therefore restrict ourselves to the abelian case. We further restrict ourselves to the case of Majorana neutrinos, which implies that there cannot be a complex eigenvalue of  the matrices $\rho(g_\nu)$ and they therefore satisfy $\rho(g_\nu)^2=1$, and  we can further choose $\det \rho(g_\nu)=1$. By further requiring three distinguishable Majorana neutrinos the group $G_\nu$ is restricted to be the Klein group $Z_2\times Z_2$. To be able to determine (up to permutations of rows and columns) the mixing matrix from the group structure it is necessary to have all neutrinos transform as inequivalent singlets of $G_\nu$. The same is true for the charged leptons which shows that $G_e$ cannot 
be smaller than $Z_3$. We can now determine the mixing via the 
unitary matrices $\Omega_e$, $\Omega_\nu$ that satisfy
\begin{align}
\Omega_e^\dagger \rho(g_e) \Omega_e= \rho(g_e)_{diag}, \qquad \Omega_\nu^\dagger \rho(g_\nu) \Omega_\nu= \rho(g_\nu)_{diag}
\end{align}
where $\rho(g)_{diag}$ are diagonal unitary matrices. These conditions determine $\Omega_e$, $\Omega_\nu$ up to a diagonal phase matrix $K_{e,\nu}$ and permutation matrices $P_{e,\nu}$
\begin{align}
\Omega_{e,\nu}\rightarrow \Omega_{e,\nu} K_{e,\nu} P_{e,\nu}.
\end{align}
It follows from Eq.~\eqref{eq:symmcond} that up to the ambiguities  of the last equation, $V_{e,\nu}$ are given by $\Omega_{e,\nu}$.
This can be seen as
$$
\Omega_e^T M_e M_e^\dagger \Omega_e^*=\Omega_e^T\rho^T M_e M_e^\dagger\rho^* \Omega_e^*=\rho_{diag}^T  \Omega_e^T M_e M_e^\dagger \Omega_e^* \rho_{diag}^*
$$
has to be diagonal (only a diagonal matrix is invariant when conjugated by an arbitrary phase matrix) and the phasing and permutation freedom can be used to bring it into the form $\diag(m_e^2,m_\mu^2,m_\tau^2)$, and analogously for $\Omega_{\nu}$. From these group theoretical considerations we can thus determine the PMNS matrix 
 \begin{align}
U_{\mathrm{PMNS}}=\Omega_e^\dagger \Omega_\nu
\end{align}
up to a permutation of rows and columns. It should not be surprising that it is not possible to uniquely pin down the mixing matrix, as it is not possible to predict lepton masses in this approach.

\begin{figure}[t]
\centering
\includegraphics[width=.9\textwidth] {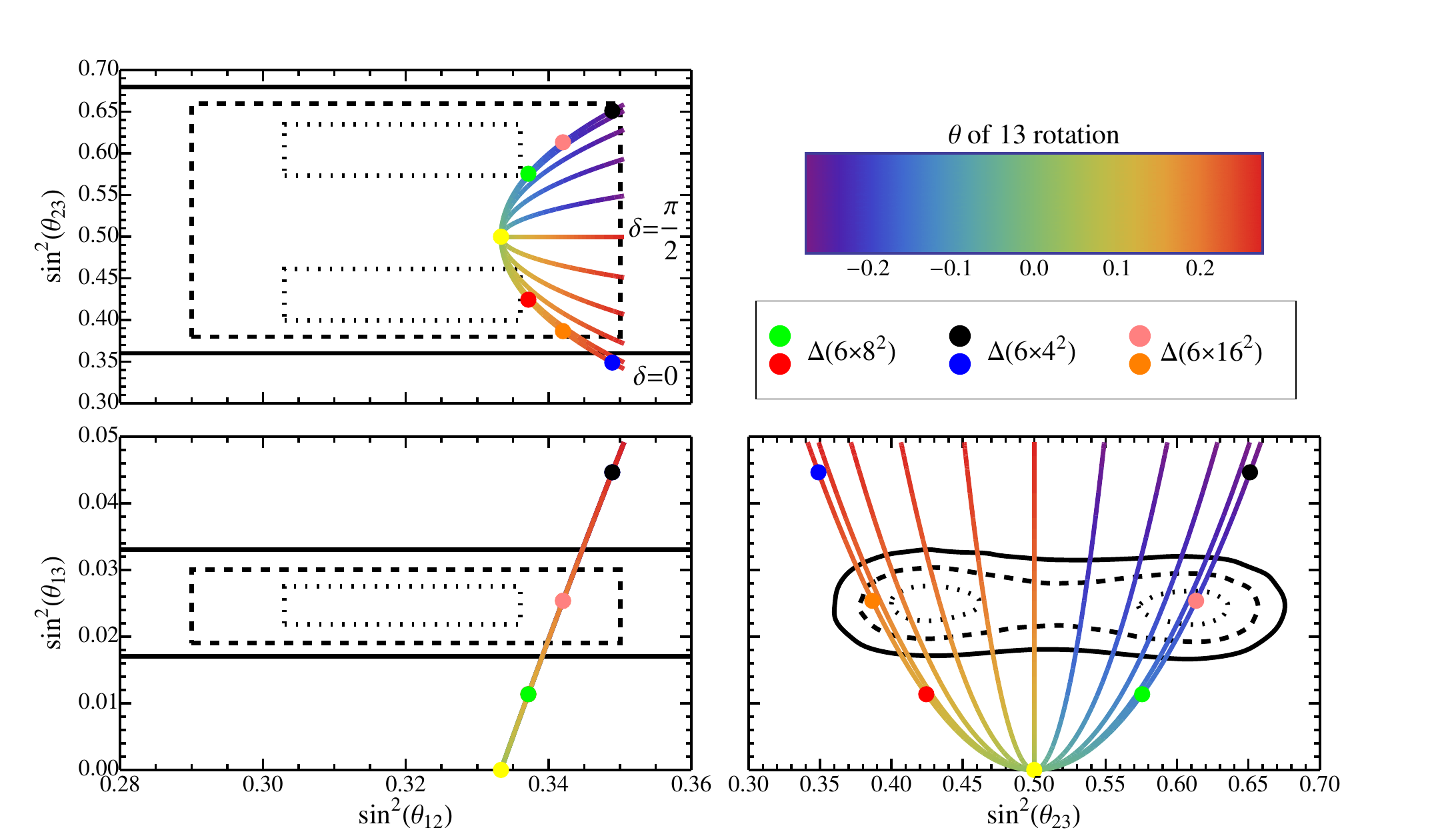} 
\caption[]{The deviations from TBM of the form $U_{\mathrm{PMNS}}=U_{\mathrm{HPS}}U_{13}(\theta, \delta)$ generated by the angle $\theta$ defined in Eq.~\eqref{eq:U13-def}. The yellow point represents TBM, the continuous lines give the deviations from TBM with the angle $\theta$ given by the color code in the top right corner for $\delta=\frac{n}{5}\frac{\pi}{2}$ for $n=0,\dots,5$, where $n=0$ is the outermost parabola etc. The global fits by Forero et al.\cite{Tortola:2012te} up to 3-sigma are depicted areas encircled by the black lines respectively. One sees that the perturbations can go in the right direction, for $\theta\sim .1-.2$ a satisfactory fit to the data can be produced. Note that the corrections to the solar angle are smaller than the corrections to the other angles. The  black and blue points are mixing patterns that can be produced by the flavor group $\Delta(96)$, the green and red points can be derived from $\Delta(384)$ while the group $\Delta(1536)$ generates mixing angles represented 
by the pink and orange points. This will be reviewed in Section \ref{sec:new-starting-points}.\label{fig:para-plot}}
\end{figure}

Let us now try to apply this machinery to some interesting cases. We have seen that the smallest residual symmetry in the charged lepton sector is given by a $G_e=\braket{T\vert T^3=E}\cong Z_3$. We use a basis where the generator is given by
\begin{align}
\rho(T)=T_3\equiv \left(\begin{array}{ccc}
0&1&0\\
0&0&1\\
1&0&0
\end{array}\right).
\label{eq:T3def}
\end{align}
This matrix will be our standard 3-dimensional representation of $Z_3$ and the notation $T_3$ will be used throughout this work. It is  diagonalized by 
\begin{align}
\Omega_e^\dagger \rho(T) \Omega_e=\diag(1,\omega^2, \omega)\qquad\mathrm{and}\qquad\Omega_e&=\Omega_T\equiv\frac{1}{\sqrt{3}}
\left(
\begin{array}{ccc}
 1 & 1 & 1 \\
 1 & \omega ^2  & \omega  \\
 1 & \omega  & \omega ^2 
\end{array}
\right),\label{eq:SigmaTdef}
\end{align}
and $\omega=e^{\I 2 \pi/3}$.
Having fixed the basis by choosing the $Z_3$ generator the way we just did, it is now essentially a question of choosing generators and studying the predicted mixing matrix. Let us first look at the case where there is only one generator $S$ of $G_\nu$, satisfying $\rho(S)^2=1$ and $\det\rho(S)=1$:
\begin{align}
\rho(S)=S_3\equiv \left(\begin{array}{ccc}
1&0&0\\
0&-1&0\\
0&0&-1
\end{array}\right).
\label{eq:S3def}
\end{align}
Due to the degenerate eigenvalues there is a two-parameter freedom in the matrix $\Omega_\nu$ and it will turn out to be useful in classifying our result later to write it as 
\begin{align}
\Omega_\nu^\dagger \rho(S) \Omega_\nu=\diag(-1,1,-1)\qquad\mathrm{with}\qquad
\Omega_\nu=\Omega_{U}U_{13}(\theta,\delta)
\end{align}
with
\begin{align}
\Omega_U
&=\left(
\begin{array}{ccc}
 0 & 1 & 0 \\
 \frac{1}{\sqrt{2}} & 0 & -\frac{i}{\sqrt{2}} \\
 \frac{1}{\sqrt{2}} & 0 & \frac{i}{\sqrt{2}}
\end{array}
\right)\qquad\mathrm{and}\qquad
U_{13}(\theta,\delta)=\left(
\begin{array}{ccc}
 \cos \theta & 0 & e^{\I \delta}\sin \theta \\
0&1&0\\
-e^{-\I \delta}\sin\theta &0&\cos \theta
\end{array}
\right).
\label{eq:U13-def}
\end{align}
Obviously this does not completely fix the leptonic mixing matrix yet, as the first and third eigenvalues are the same and the corresponding eigenstates can be rotated into each other without breaking the symmetry. To completely fix the mixing matrix we have to enlarge $G_\nu$ by another generator. Let us look at the effect of adding the symmetry generator $U$ with
\begin{align}
\rho(U)=U_3\equiv -\left(\begin{array}{ccc}
1&0&0\\
0&0&1\\
0&1&0
\end{array}\right).
\label{eq:U3-def}
\end{align}
 This fixes the value of $\theta$ to zero, 
$
\Omega_U^\dagger \rho(U) \Omega_U=\diag(-1,-1,1),
$
 and thus the mixing matrix to the famous tri-bimaximal mixing (TBM) form
\begin{align}
U_{\mathrm{PMNS}}=\Omega_e^\dagger \Omega_U=U_{\mathrm{HPS}}\equiv\left(
\begin{array}{ccc}
\sqrt{\frac{2}{3}}&\frac{1}{\sqrt{3}}&0\\
-\frac{1}{\sqrt{6}}&\frac{1}{\sqrt{3}}&\frac{1}{\sqrt{2}}\\
-\frac{1}{\sqrt{6}}&\frac{1}{\sqrt{3}}&-\frac{1}{\sqrt{2}} 
\end{array}\right),
\label{eq:HPS-matrix}
\end{align} 
which corresponds to the mixing angles $\sin^2\theta_{12}=\frac13$ , $ \sin^2\theta_{23}=\frac12$,and $\sin^2\theta_{13}=0$. TBM pattern is predicted by the discrete group $S_4=\langle S_3, T_3, U_3 \rangle$ \cite{Lam:2007fk,Lam:2008fj,Lam:2008sh} with $S_3$, $T_3$ and $U_3$ given in our example above. Until very recently, this pattern gave a good description of the mixing matrix and the fact that this mixing pattern can be obtained from simple symmetry considerations has prompted a lot of model building activity (for a general scan of models based on flavor groups up to the order of 100, see \cite{Parattu:2010cy}). In light of the recent measurement of a non-vanishing $\theta_{13}$ there has been interest in the physical situation where the (broken) flavor symmetry does not fully determine the mixing angles. For example if the residual symmetry in the neutrino sector $G_\nu $ is taken to be $G_\nu=\vev{S}\cong Z_2$, we have seen that the leptonic mixing matrix is given by $U_{\mathrm{PMNS}}=U_{\mathrm{HPS}}
U_{13}(\theta, \delta)$ \cite{Grimus:2008ve,Albright:2008rp,Ge:2011qn,Ge:2011ih,Hernandez:2012ra,Feruglio:2012cw,King:2013eh}. This mixing pattern leads to $\sin^2 \theta_{12}>1/3$ and is sometimes known as tri-maximal pattern TM2. The result of a deviation from $\theta=0$ in terms of mixing angles is shown in Figure \ref{fig:para-plot} where one can read off that a $13$-rotation about an angle of $\theta\simeq 0.2$ is required. In Section \ref{sec:new-starting-points}, we will show that all flavor groups with the order less than 1536 with an additional group $\Delta(6\cdot 16^2)$ lie on the parabola $U_{\mathrm{PMNS}}=U_{\mathrm{HPS}}U_{13}(\theta, \delta=0)$.

It should be clear from the discussion above that, a different choice of generators of the residual symmetry groups leads to different mixing pattern. For example, if we take the $G_\nu$ not to be the Klein group $G_\nu=\braket{S,U\vert S^2=U^2=E;SU=US}\cong Z_2\times Z_2$ that leads to TBM, but the isomorphic group $G_\nu=\braket{S,X\vert S^2=X^2=E;SX=XS}\cong Z_2\times Z_2$ with $X=T^2ST$ and 
 \begin{align}
\rho(X)=\left(\begin{array}{ccc}
-1&0&0\\
0&1&0\\
0&0&-1
\end{array}\right)
\end{align}
fixes 
\begin{align}
\Omega_X^\dagger \rho(X) \Omega_X=\diag(-1,-1,1)\qquad\mathrm{with}\qquad
\Omega_X=\Omega_{U}U_{13}(\theta=-\frac{\pi}{4}, \delta=\frac{\pi}{2})
\end{align}
which gives 
\begin{align}
\parallel U_{\mathrm{PMNS}} \parallel =\parallel \Omega_e^\dagger \Omega_X \parallel=\frac{1}{\sqrt{3}}\left(
\begin{array}{ccc}
1&1&1\\
1&1&1\\
1&1&1
\end{array}\right),
\end{align} 
which corresponds to the mixing angles $\sin^2\theta_{12}=\frac 12\   , \sin^2\theta_{13}=\frac 13 \  \mathrm{and}\  \sin^2\theta_{23}=\frac 12$. This mixing pattern is sometimes called tri-maximal mixing \cite{Cabibbo:1977nk,Wolfenstein:1978uw,Giunti:1994mh,Harrison:1999cf}.

\section{New Starting Points}
\label{sec:new-starting-points}
In Section \ref{sec:lepton-mixing-finite-groups} we have seen that to uniquely (up to permutations of rows and columns) determine the mixing patterns from group theoretic considerations, it is essential to have unbroken remnant symmetries in the charged lepton and neutrino sectors and to have enough symmetries in each sector that there are three inequivalent representations in each sector. For the neutrino sector one needs a Klein group $Z_2\times Z_2$ for three generations of Majorana neutrinos as there cannot be a cyclic group of order larger than 2 in the neutrino sector as complex representations would forbid neutrino mass terms. Higher $Z_2^n$ ($n>2$) product groups are redundant as the Klein group is the maximal group that one could have for three generations of neutrinos. The condition of $G_\nu=Z_2$ \cite{Ge:2011qn,Ge:2011ih,Hernandez:2012ra,Feruglio:2012cw,Hernandez:2012xx,He:2011kn,Ge:2010js} can be used to determine the mixing angles up to a two-parameter freedom. This approach however does not 
allow a sharp 
mixing angles prediction but can accommodate the experimental results more easily. A group scan based on this assumption appeared in \cite{Lam:2012he}. For the charged lepton sector we first consider a cyclic group with a $Z_3$ symmetry, and generalize it later to general finite abelian groups. 

\subsection{\mbox{\texorpdfstring{$G_{\nu}=Z_2\times Z_2$, $G_e=Z_3$}{Ge=Z3}}}\label{sec:z3}
We first assume that the charged lepton sector has a $Z_3$ symmetry. The way to generate new mixing structures apart from TBM is now to embed these abelian symmetries in different ways into some larger group. Once the group and the embedding is specified, the Clebsch-Gordon coefficients of the specified group uniquely (up to permutation of rows and columns) specify the mixing pattern.

\begin{table}
\centering
\begin{tabular}{llcllcllcll}\toprule
 $n$ & $G$ & & $n$ &$ G$ & &$n$  &$G$ & &$n$  &$G$ \\ 
 \cmidrule{1-2}\cmidrule{4-5}\cmidrule{7-8}\cmidrule{10-11}
$4$& $\Delta(6\cdot 4^2)$ & & $9$& $(Z_{18}\times Z_6)\rtimes S_3$  &    &$13$& $\Delta(6\cdot 26^2)$&  &  $18$& $(Z_{18}\times Z_6)\rtimes S_3$   \\  
$5$& $\Delta(6\cdot 10^2)$ &  &$10$& $\Delta(6\cdot 10^2)$  &  & $14$& $\Delta(6\cdot 14^2)$   & &$24$& $Z_3 \times \Delta(6\cdot 8^2)$  \\  
$7$& $\Delta(6\cdot 14^2)$ &  &$11$& $\Delta(6\cdot 22^2)$ & &   $15$& $Z_3\times \Delta(6\cdot 10^2)$  &  &  & \\  
$8$& $\Delta(6\cdot 8^2)$ &  &$12$& $Z_3\times\Delta(6\cdot 4^2)$  &   & $16$& $\Delta(6\cdot 16^2)$  &   &  &\\ \bottomrule
\end{tabular} 
\caption{Groups generated by $T_3$, $S_3$ and $U_3(n)$, that lead to new starting points. The series of groups $\Delta(6 n^2)$ has been studied in \cite{Escobar:2009mz} and the group $(Z_{18}\times Z_6)\rtimes S_3$, apart from being defined by $\vev{T_3,S_3,U_3(9)}$, is the group number 259 of order 648 in the \SmallGroups\ catalogue of \GAP~\cite{SmallGroups:2011}.}
\label{tab:startingpoint-deltas}
\end{table}
\ra{1}

\begin{figure}\centering
\includegraphics[width=1\textwidth] {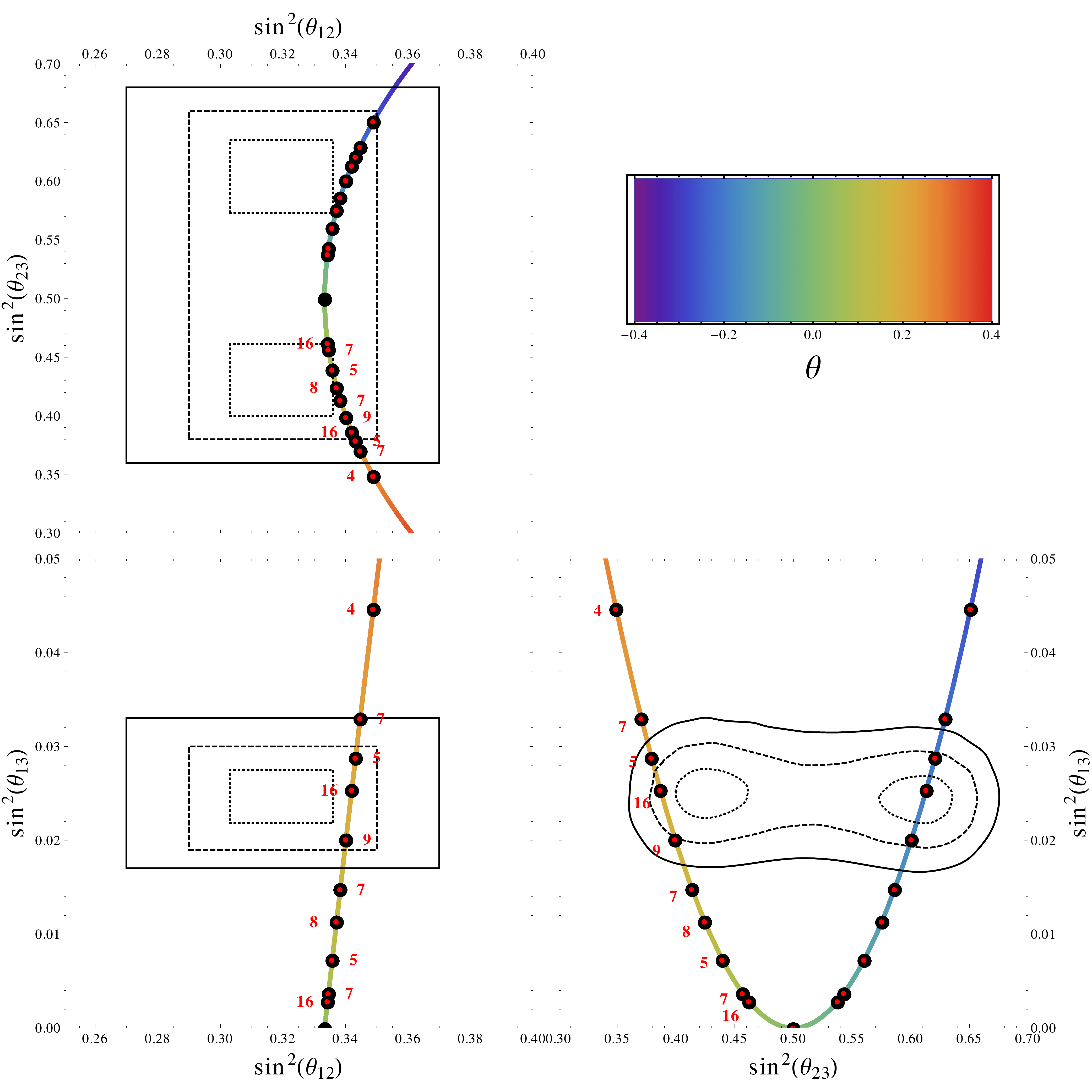} 
\caption[]{The leptonic mixing angles (black circles) determined from our group scan up to order 1536 are shown. The red dots represent the mixing angles that we have determined from the generator $S_3$, $T_3$ and $U_3(n)$. The red labels represent the integer $n$ that generates the $U_3(n)$ matrix. The interpolating line is colored according to the value of $\theta$ as defined in Eq.~\eqref{eq:U13-def}. See the main text for more detailed informations. We have also omitted the labeling of larger $n$ that generates the same repeating groups or mixing angles.}
\label{fig:starting-points}
\end{figure}

\begin{figure}\centering
\includegraphics[width=1\textwidth]{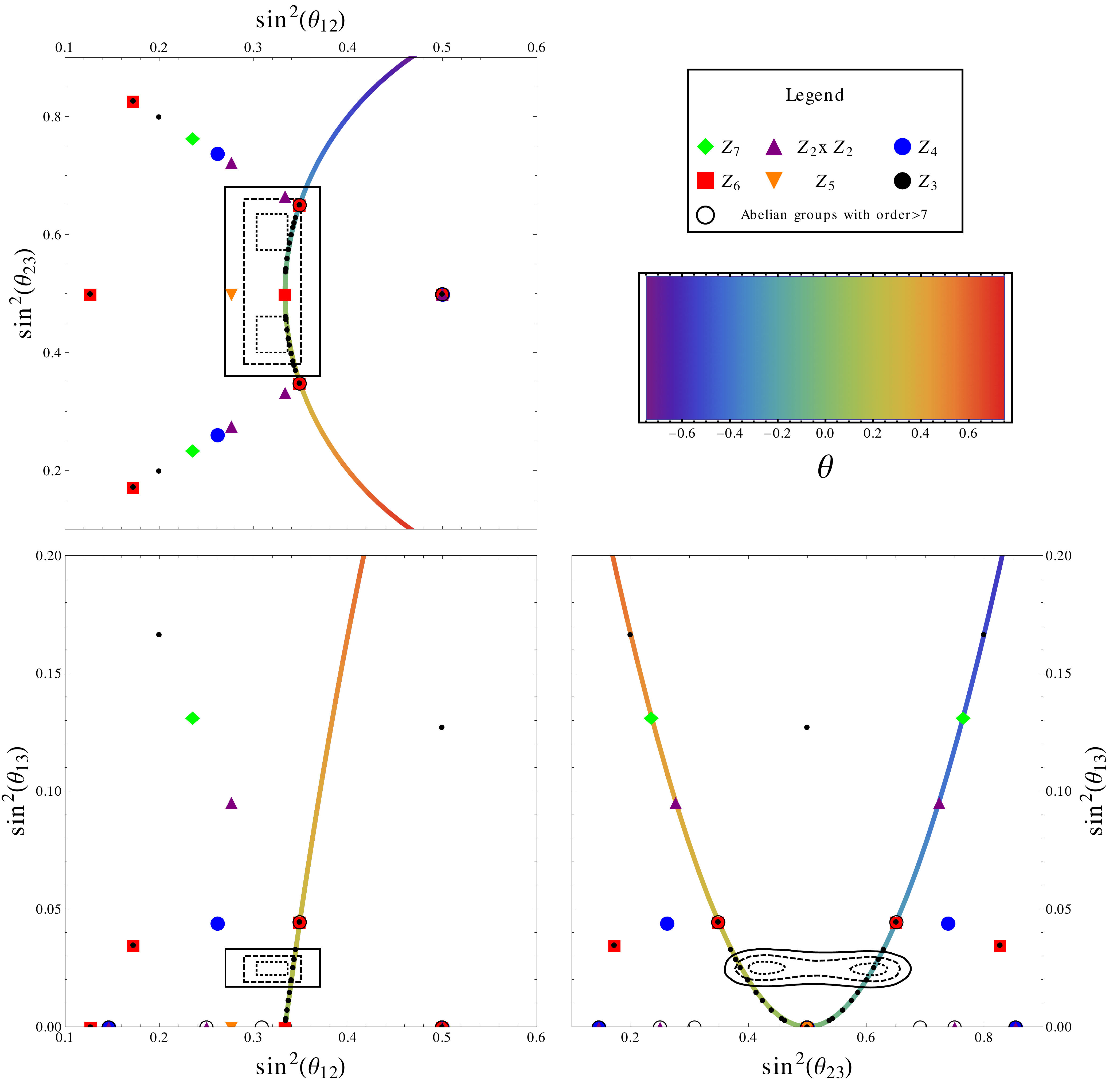} 
\caption{Mixing angles obtained from groups with all the abelian subgroups of $G_e$ up to order 511. For comparison purpose we also plotted the mixing patterns obtained from groups with $G_e=Z_3$. Only the mixing patterns generated by finite modular groups and their subgroups are obtained for $|G_e|>3$, see Ref.~\cite{de-Adelhart-Toorop:2012fv}. The mixing patterns from $G_e=Z_3$ that lie outside the parabola are also generated by finite modular groups (e.g. $A_5$, $PSL(2,Z_7)$) and their subgroups.}
\label{fig:abelian}
\end{figure}

We have performed such an analysis, considering all discrete groups of size smaller than 1536 using the computer algebra program \GAP. The total number of 1336749 finite discrete groups (with an additional group of $\Delta(6\cdot 16^2)$) has been scanned, keeping only groups which contain the Klein group and $Z_3$ as their subgroups. To speed up the scanning process, we use the Lagrange theorem to skip over groups with order that is not divisible by 4 (order of Klein group) and 3 (order of $Z_3$). We then look for groups that carry at least a 3-dimensional irreducible representation. Furthermore we only consider 3-dimensional representations of a specific group if they are faithful to avoid duplicating subgroups which have been scanned. All the 3-dimensional representations $\rho(U)$, $\rho(S)$ and $\rho(T)$ for generators $U$, $S$ and $T$ are then being recorded. Note that one should not be confused by $\rho(S)$, $\rho(U)$ and $\rho(T)$ with $S_3$, $U_3$ and $T_3$ as $S_3$, $U_3$ and 
$T_3$ are specific matrices defined in previous section. With each set of generators for the Klein group, one can determine the unitary matrix $\Omega_{\nu}$ which simultaneously diagonalizes both $\rho(S)$ and $\rho(U)$ of the Klein group. The corresponding $\Omega_{\nu}$ will be subsequently multiplied with a unitary matrix $\Omega_e$ that diagonalizes the $\rho(T)$ matrix in the same representation such that a PMNS matrix can be obtained. This process is repeated for all the different combinations of $\Omega_{\nu}$ and $\Omega_e$ obtained from different sets of $\{\rho(U), \rho(S) \}$ and $\rho(T)$ respectively. All the matrices with permutations of rows and columns for the corresponding PMNS matrices were subsequently generated and we ordered the PMNS matrices with the smallest 13-entry and an additional condition such that the 11-entry is larger or equal to 12-entry. A huge amount of duplicates is removed with this method. The flavor mixing angles are extracted from the 
remaining PMNS matrices and the results are plotted as black circles in Figure \ref{fig:starting-points}. From Figure \ref{fig:starting-points} we observe that from about one million finite discrete groups that we have scanned, only 3 finite groups up to the order of 1535 including $\Delta(6\cdot 16^2)$ generate the mixing angles which are allowed by the experimental values up to $3\sigma$, with the strongest constraint coming from $\theta_{13}$. Amazingly many of the discrete finite groups with $G_e=Z_3$ and $G_{\nu}=Z_2\times Z_2$ that we have scanned yield a combination of leptonic mixing angles that lie on a parabola. However in general there exist also mixing angles that lie beyond our plotted region, these points will be shown in Figure \ref{fig:abelian}. 

Those discrete groups that lie on the parabola can be easily presented in a simple systematic way:  in Section \ref{sec:lepton-mixing-finite-groups}, we had seen that if the $Z_3$ symmetry in the charged lepton sector is generated by the matrix $T_3$ given in Eq.~\eqref{eq:T3def} and the Klein group is generated by the matrices $S_3$ and $U_3$ of Eqs.~\eqref{eq:S3def} and \eqref{eq:U3-def}, the resulting mixing matrix is of the TBM form Eq.~\eqref{eq:HPS-matrix}. Now, all new mixing patterns found in the scan can be written as 
\begin{align}
U_{\mathrm{PMNS}}=U_{\mathrm{HPS}}U_{13}(\theta=\frac{1}{2} \arg(z), \delta=0)
\label{eq:final}
\end{align}
which is the result of the remnant symmetry $T_3$ in the charged lepton sector and the choice of $S_3$ and
 \begin{align}
 U_3(n) =-\left( \begin{array}{ccc} 1&0&0\\0&0&z\\0&z^*&0 \end{array}\right) \qquad \mathrm{with}\qquad \ev{z}\cong Z_n,\;   n\in \mathbb{N}
 \label{eq:Un-def}
 \end{align}
as generators of a Klein group in the neutrino sector. One can think of $U_3(n)$ as one of the more generalized generator for the Klein group. Note that for any $z$ with modulus one, we have $[S_3,U_3(n) ]=0$ and $U_3(n)^2=\mathbbm{1}_3$ and therefore the group generated by $S_3$ and $U_3(n)$ is always a Klein group, $\ev{S_3,U_3(n)}\cong Z_2\times Z_2$. For the group generated by $T_3$, $S_3$ and $U_3(n)$ to be finite, $z$ has to be of the form given in Eq.~\eqref{eq:Un-def}, as may be seen by looking at the group element $(U_3(n) T_3)^2=\diag(z,z,{z^*}^2)$ which is of finite order $n \in \mathbb{N}$ iff $z^n=1$. The requirement  $\ev{z}\cong Z_n$ further fixes $n$ to be the smallest $n $ for which $z^n=1$. Note that different $n$-th root of $z$ will in general lead to different leptonic mixing angles via Eq. (\ref{eq:final}), as can be seen in Figure 
\ref{fig:starting-points}. The group generated by $T_3$, $S_3$ and $U_3(n)$ is always the same due to the requirement $\ev{z}\cong Z_n$. The names of the groups generated for $n=4,\dots,16$ can be found in Table \ref{tab:startingpoint-deltas} and the groups $\Delta(96)$ ($n=4$) and $\Delta(384)$ ($n=8$) have been obtained before in \cite{de-Adelhart-Toorop:2012fv}. Note that not all of the groups generated in this way can be classified as $\Delta(6\cdot n^2)$, e.g. the group $(Z_{18}\times Z_6)\rtimes S_3$\footnote{Let us mention that $(Z_{18}\times Z_6)\rtimes S_3$ is a subgroup of $\Delta(6\cdot 18^2)$, however we only consider the smallest group that generates a given mixing pattern.}. Another surprising observation is that all the groups which are restricted to those shown in Figure \ref{fig:starting-points} give a prediction of $\delta_{\mathrm{CP}}=\pi$ or $\delta_{\mathrm{CP}}=0$. In general other groups that we have scanned which lie outside the region shown in Figure \ref{fig:starting-points} do 
give non-trivial Dirac CP phases.
\begin{table}
 \centering
 \begin{tabular}{|c|c|c|ccc|}
\hline
$n$&$G$ & \GAP-Id & $\sin^2(\theta_{12})$ & $\sin^2(\theta_{13})$ & $\sin^2(\theta_{23})$ \\
\hline 5 &$\Delta(6\cdot 10^2)$ & $[600,179]$ & $0.3432$ & $0.0288$ & $0.3791$ \\
&&&$0.3432$ & $0.0288$ & $0.6209$ \\
9 &$(Z_{18}\times Z_6)\rtimes S_3$ & $[648,259]$ & $0.3402$ & $0.0201$ & $0.3992$ \\
&&& $0.3402$ & $0.0201$ & $0.6008$ \\
16& $\Delta(6\cdot 16^2)$ & n.a. & $0.3420$ & $0.0254$ & $0.3867$ \\
&&& $0.3420$ & $0.0254$ & $0.6134$ \\ \hline
\end{tabular}
\caption{Mixing angles which are compatible with experimental results generated by flavor groups up to order 1536. The group identification function in \SmallGroups\ is not available for group with order 1536.}
\label{tab:mixingangles}
\end{table}
The predictions for mixing angles for all groups of size smaller or equal than 1536 groups is presented in Figure \ref{fig:starting-points}. It should be clear that if one allows for groups of arbitrary size, the parabola depicted in Figure \ref{fig:starting-points} will be densely covered. As one can see, the mixing patterns corresponding to $n=5$, $n=9$ and $n=16$ give a good descriptions of the leptonic mixing matrix. See Table  \ref{tab:mixingangles} for the resulting mixing angles.

\subsection{\mbox{\texorpdfstring{$G_{\nu}=Z_2\times Z_2$, $|G_e|>3$}{Geabelian}} }
In this section we will discuss the result of all the neutrino mixing angles scanned by relaxing the condition on the charged lepton subgroups. We allow $G_e$ to be any abelian groups which are the subgroups of the original flavor group $G_f$. The Klein group is kept as a subgroup of the remnant symmetry for the neutrino mass matrix. The scanning procedure is performed as in Section \ref{sec:z3}, with a few subtleties changed. For instance if $G_e$ consists of two or more generators above, the unitary matrix $\Omega_e$ is constructed as a matrix which diagonalizes all the generators of $G_e$ simultaneously. We have scanned all the discrete finite groups $G_f$ up to order 511 with all the abelian subgroups $G_e$. The result is shown in Figure \ref{fig:abelian} and we obtain only the neutrino mixing angles predicted by finite modular groups and their subgroups, as discussed in 
\cite{de-Adelhart-Toorop:2012fv}. Mixing angles which are not generated by finite modular groups appear however if we scan the groups with order higher than 511. For instance in Section \ref{sec:z3} we have seen groups such as $\Delta(6\cdot10^2)$ and $(Z_{18}\times Z_6)\rtimes S_3$ that are not generated by finite modular groups but give a set of mixing angles which are in good agreement with experiments.  

\section{Conclusions}
\label{sec:conclusion}
We have presented an extensive scan over discrete symmetry groups that may be used to predict leptonic mixing angles. It is assumed that the left-handed leptonic doublets are assigned to 3-dimensional representations and that the symmetry is broken to different mismatched symmetry groups $G_e$ and $G_\nu$ in the charged lepton and neutrino sectors, respectively. With the assumptions of $G_e\cong Z_3$ and $G_\nu\cong Z_2 \times Z_2$, we have scanned all groups with size less than 1536 with the additional group $\Delta(6\cdot 16^2)$ and surprisingly we found only 3 groups that lead to acceptable mixing patterns. In particular the groups $\Delta(6\cdot 10^2)$, $(Z_{18}\times Z_6)\rtimes S_3$ and $\Delta(6\cdot 16^2)$ were found to predict the leptonic mixing angles within 3-sigma region of global fits. In all the interesting cases we predicted $\delta_{\mathrm{CP}}=0$ or $\delta_{\mathrm{CP}}=\pi$. These groups should be interesting for model building and phenomenological studies.

We have also found a way to systematically categorize groups that generate the leptonic mixing angles which lie on the parabola that can be described by a 13-rotation of TBM mixing matrix, i.e. Eq.~\eqref{eq:final}. It is possible that if our result can be extrapolated to arbitrary large order of discrete finite groups, with the assumption that the remnant symmetries of $G_e$ and $G_\nu$ are not broken, all of the interesting groups should generate lepton mixing patterns that lie on the parabola with trivial Dirac CP phase.

In a second scan we relaxed the condition $G_e\cong Z_3$ and assumed $G_e$ to be any abelian group. In this scan up to groups of size 511 no new interesting groups were found.

From all the discrete groups that we have scanned, only groups with relatively large order predict acceptable mixing angles. This might seem not economical from the model building perspective. However it is possible that some of the remnant symmetries are accidental, e.g. in the case of $A_4$ models that predict TBM, the generator $U_3$ is not part of the group $A_4$ but rather represents an outer automorphism of the group $A_4$ \cite{Holthausen:2012dk}. It is possible that there exist some smaller groups whose automorphism group is isomorphic to the groups we have found (for example the smallest group whose automorphism group contains $\Delta(96)$ is of the order of 32 \cite{Holthausen:2012dk}). This idea is interesting from group theoretical and phenomenological perspective and such an investigation is left for future studies. 
\\[10pt]
\textbf{Acknowledgements:} M.H. and K.S.L. acknowledge support by the International Max Planck Research School for Precision Tests of Fundamental Symmetries.

\bibliography{biblio2}

\end{document}